\documentstyle[preprint,aps]{revtex}
\preprint{USM-TH-138}
\begin{document}
\title{Static potential and a new generalized connection in three dimensions}
\author{ Patricio Gaete \thanks{E-mail: patricio.gaete@fis.utfsm.cl}}
\address{Departamento de F\'{\i}sica, Universidad T\'ecnica F.
Santa Maria, Valpara\'{\i}so, Chile} \maketitle

\begin{abstract}
For a recently proposed pure gauge theory in three dimensions,
without a Chern-Simons term, we calculate the static interaction
potential within the structure of the gauge-invariant variables
formalism. The result coincides with that of the
Maxwell-Chern-Simons theory in the short distance regime, which
shows the confining nature of the potential.
\end{abstract}
\smallskip

PACS number(s): 11.10.Ef, 11.10.Kk

\section{INTRODUCTION}

Systems in (2+1) dimensions have been extensively discussed in the
last few years\cite{Deser,Dunne,Khare}. As is well known, the
interest in studying three-dimensional theories is mainly due to
the possibility of realizing fractional statistics, where the
physical excitations obeying it are called anyons. The
three-dimensional Chern-Simons gauge theory is the key example, so
that Wilczek's charge-flux composite model of the anyon can be
implemented\cite{Wilczek}. In this context it may be recalled that
when the Chern-Simons term is added to the usual Maxwell term, the
gauge field becomes massive, leading to topologically massive
gauge theories. Interestingly, the statistics of anyons changes
with their mutual distance in the presence of the Maxwell term
\cite{Shisuya}. We further note that recently a novel way to
describe anyons has been considered \cite{Itzhaki}. The crucial
ingredient of this development is to introduce a generalized
connection in (2+1) dimensions which permits to realize fractional
statistics, such that the Chern-Simons term needs not be
introduced. More precisely, it was shown that a Lagrangian which
describes Maxwell theory coupled to the current via the
generalized connection leads to fractional statistics by the same
mechanism, as in the case of the Maxwell-Chern-Simons theory, of
attaching a magnetic flux to the electrons.

On the other hand, we also recall that the ideas of screening and
confinement play a central role in gauge theory. In this
connection the interaction potential between static charges is an
object of considerable interest, and its physical content can be
understood when a correct separation of the physical degrees of
freedom is made.

With this in view, we will apply in this work a formalism in terms
of physical (gauge-invariant) quantities. This method was used
previously for studying features of screening and confinement in
two-dimensional quantum electrodynamics, generalized
Maxwell-Chern-Simons gauge theory and for the Yang-Mills field
\cite{Gaete}. This method is applied below to study the
interaction energy in the recently proposed three-dimensional
electrodynamics \cite{Itzhaki}. The procedure exploits the rich
structure of the dressing around static fermions, where we refer
to the cloud made out of the vector potentials around the fermions
as dressing. The methodology presented here provides a
physically-based alternative to the usual Wilson loop approach.

\section{INTERACTION ENERGY}

As already stated, our objective is to compute explicitly the
interaction energy between static pointlike sources for this new
electrodynamics. For this purpose we shall first carry out its
Hamiltonian analysis. The gauge theory we are considering is
defined by the following Lagrangian in three-dimensional
space-time:
\begin{equation}
{\cal L} =  - \frac{1}{4}F_{\mu \nu }^2  - A_\mu ^\theta  J^\mu =
- \frac{1}{4}F_{\mu \nu }^2  + \frac{\theta }{2}\varepsilon _{\mu
\nu \rho } J^\mu  F^{\nu \rho }  - A_\mu  J^\mu. \label{gen1}
\end{equation}
Here $ A_\mu ^\theta   = A_\mu   - \frac{\theta }{2}\varepsilon
_{\mu \nu \rho } F^{\nu \rho }$ is the generalized connection, $
J_\mu$ is the external current and ${\theta}$ is a parameter with
dimension $ M^{ - 1}$. The canonical momenta are $ \Pi ^\mu   =
F^{\mu 0}  + \theta \varepsilon ^{0\mu \nu } J_\nu$, which results
in the usual primary constraint $\Pi ^0=0$ and $\Pi ^i  = F^{i0} +
\theta \varepsilon ^{ij} J_j$ ($i,j=1,2$ ). The canonical
Hamiltonian is given by
\begin{equation}
H_C  = \int {d^2 x} \left\{ { - \frac{1}{2}F_{i0} F^{i0}  +
\frac{1}{4}F_{ij} F^{ij}  - A_0 \left( {\partial _i \Pi ^i  - J^0
} \right) - \frac{\theta }{2}\varepsilon _{ij} J^0 F^{ij}  + A_i
J^i } \right\}, \label{gen2}
\end{equation}
and it is straightforward to see that the preservation in time of
the primary constraint leads to the secondary constraint
\begin{equation}
\Omega _1 \left( x \right) \equiv \partial _i \Pi ^i \left( x
\right) - J^0 \left( x \right)=0. \label{Gauss}
\end{equation}
The above constraints are the first-class constraints of the
theory since no more constraints are generated by the time
preservation of the secondary constraint (\ref{Gauss}). The
corresponding total (first class) Hamiltonian that generates the
time evolution of the dynamical variables then reads
\begin{equation}
H = H_C  + \int {d^2 x} \left\{ {c_0 \left( x \right)\Pi _0 \left(
x \right) + c_1 \left( x \right)\Omega _1 \left( x \right)}
\right\}, \label{gen3}
\end{equation}
where $ c_0 \left( x \right)$ and  $ c_1 \left( x \right)$ are the
Lagrange multiplier fields to implement the constraints. Since $
\Pi ^0 =0$ for all time and  $ \dot{A}_0 \left( x \right) = \left[
{A_0 \left( x \right),H} \right] = c_0 \left( x \right)$, which is
completely arbitrary, we eliminate $ A^0 $ and $ \Pi^0 $ because
they add nothing to the description of the system. Thus the
Hamiltonian takes the form
\begin{equation}
H = \int {d^2 x} \left\{ { - \frac{1}{2}F_{i0} F^{i0}  +
\frac{1}{4}F_{ij} F^{ij}  + c^\prime  \left( x \right)\left(
{\partial _i \Pi ^i  - J^0 } \right) - \frac{\theta
}{2}\varepsilon _{ij} J^0 F^{ij}  + A_i J^i } \right\}
\label{gen4}
\end{equation}
where $ c^{\prime }(x)=c_{1}(x)-A_{0}(x)$.

In accordance with the Dirac method, we impose one gauge
constraint such that the full set of constraints becomes second
class. A convenient choice is found to be
\begin{equation}
\Omega _2 \left( x \right) \equiv \int\limits_{C_{\xi x} } {dz^\nu
} A_\nu  \left( z \right) = \int_0^1 {d\lambda } x^i A_i \left(
{\lambda x} \right) = 0 \label {gen5}
\end{equation}
where  $\lambda$  $\left( {0 \le \lambda  \le 1} \right)$ is the
parameter describing the spacelike straight path between the
reference point $ \xi ^k $ and $ x^k $ , on a fixed time slice.
For simplicity we have assumed the reference point $\xi^k=0$. The
choice (\ref{gen5}) leads to the Poincar\'{e} gauge \cite{Gaete}.
Through this procedure, we arrive at the following set of Dirac
brackets for the canonical variables
\begin{equation}
\left\{ A_{i}(x),A^{j}(y)\right\} ^{*}=0,  \label{gen6}
\end{equation}
\begin{equation}
\left\{ \pi _{i}(x),\pi ^{j}(y)\right\} ^{*}=0,\label{gen6b}
\end{equation}
\begin{equation}
\left\{ A_{i}(x),\pi ^{j}(y)\right\} ^{*}=g _{i}^{j}\delta
^{(2)}\left(
x-y\right) -\partial _{i}^{x}\int_{0}^{1}d\lambda \text{ }%
x^{j}\delta ^{(2)}\left( \lambda x-y\right) .  \label{gen7}
\end{equation}
In order to illustrate the discussion, we now write the Dirac
brackets in terms of the magnetic $\left( B=\varepsilon
_{ij}\partial ^{i}A^{j}\right)$ and electric $( E^i  = \Pi ^i  -
\theta \varepsilon ^{ij} J_j)$ fields as
\begin{equation}
\left\{ {E_i \left(  x \right),E_j \left(  y \right)} \right\}^
* = 0 , \label{gen8}
\end{equation}
\begin{equation}
\left\{ {B \left(  x \right),B \left(  y \right)} \right\}^
* = 0 , \label{gen9}
\end{equation}
\begin{equation}
\left\{ {E_i \left( x \right),B\left(  y \right)} \right\}^
* = - \varepsilon _{ij} \partial _x^j \delta ^{(2)} \left( x - y
\right). \label{gen10}
\end{equation}
It is important to realize that, unlike the Maxwell-Chern-Simons
theory, in the present model, the brackets (\ref{gen6}) and
(\ref{gen8}) are commutative. One can now easily derive the
equations of motion for the electric and magnetic fields. We find
\begin{equation}
 {\dot E}_i( x) =  - \varepsilon _{ij}
\partial ^j B\left(  x \right) + \theta \varepsilon _{ij} \partial
^j J^0 \left(  x \right) + J_i \left(  x \right) + \int {d^2 y}
J^k \left(  y \right)\partial _k \int\limits_0^1 {d\lambda } y_i
\delta ^{(2)} \left( {\lambda  x -  y} \right), \label{gen11}
\end{equation}
\begin{equation}
{\dot B}\left(  x \right) =  - \varepsilon _{ij} \partial _i E_j
\left(  x \right). \label{gen12}
\end{equation}
In the same way, we write the Gauss law as:
\begin{equation}
\partial _i E_L^i  + \theta \varepsilon ^{ij} \partial _i J_j  - J^0  =
0, \label{gen13}
\end{equation}
where $E_L^i$ refers to the longitudinal part of $E^i$. This
implies that for a static charge located at $x^i=0$, and $J^i=0$,
the static electromagnetic fields are given by
\begin{equation}
B\left( x \right) = \theta J^0, \label{ecua1}
\end{equation}
\begin{equation}
E_i \left( x \right) =  - \frac{{\partial _i J^0 }}{{\nabla ^2 }}
, \label{ecua2}
\end{equation}
where ${\bf \nabla} ^2$ is the two-dimensional Laplacian. For $
J^0 \left( {t,x} \right) = e\delta ^2 \left( x \right)$,
expressions (\ref{ecua1}) and (\ref{ecua2}) become
\begin{equation}
B\left( x \right) = e\theta \delta ^{\left( 2 \right)} \left( x
\right), \label{ecua3}
\end{equation}
\begin{equation}
E_i \left( x \right) =  - \frac{e}{{2\pi }}\frac{{x^i }}{{r^2 }},
\label{ecua4}
\end{equation}
with $ r \equiv |\bf x|$. One immediately sees that the associated
magnetic field has its support only at the position of the charge,
and a long range electric field is also generated. As a
consequence, the total magnetic flux associated to the magnetic
field is
\begin{equation}
\Phi _B  = \int\limits_V {d^2 } xB\left( x \right) = e\theta .
\label{ecua5}
\end{equation}
This tells us that the charged particle actually behaves like a
magnetic flux point. Accordingly, the mechanism of attaching a
magnetic flux to the charges has been implemented in a
particularly simple way, as was claimed in Ref. \cite{Itzhaki}.

After achieving the quantization we may now proceed to calculate
the interaction energy between pointlike sources in the model
under consideration. To do this, we will compute the expectation
value of the energy operator $H$ in a physical state $ \left|
\Omega \right\rangle$. We also recall that the physical states $
\left| \Omega  \right\rangle$ are gauge-invariant \cite{Dirac}. In
that case we consider the stringy gauge-invariant $\left|
{\overline \Psi  \left( \bf y \right)\Psi \left( {\bf y^ \prime }
\right)} \right\rangle$ state,
\begin{equation}
\left| \Omega  \right\rangle  \equiv \left|\overline \Psi \left(
{\bf y} \right)\Psi \left( {\bf y^\prime} \right)   \right\rangle
= \overline \psi \left( {\bf y} \right)\exp \left(
{-ie\int\limits_{\bf y}^{\bf y^\prime} {dz^i A_i \left( z \right)}
} \right)\psi \left( {\bf y^\prime} \right)\left| 0 \right\rangle,
\label{est}
\end{equation}
where $\left| 0 \right\rangle$ is the physical vacuum state and
the integral is to be over the linear spacelike path starting at
$\bf y$ and ending at $\bf y^ \prime$, on a fixed time slice. Note
that the strings between fermions have been introduced to have a
gauge-invariant state $ \left| \Omega  \right\rangle$, in other
terms, this means that the fermions are now dressed by a cloud of
gauge fields. We can write the expectation value of $H$ as
\begin{equation}
\left\langle H \right\rangle _\Omega   = \left\langle \Omega
\right|\int {d^2 } x\left( {\frac{1}{2}E_i^2  + \frac{1}{2}B^2  -
\theta BJ^0  + J^i A_i } \right)\left| \Omega  \right\rangle.
\label{pot1}
\end{equation}
The preceding Hamiltonian structure thus leads to the following
result
\begin{equation}
\left\langle H \right\rangle _\Omega   = \left\langle H
\right\rangle _0  +  \ \frac{{e^2 }}{2}\int {d^2 x\left(
{\int\limits_{\bf y }^{\bf y^\prime} {dz_i \delta ^{(2)} \left( {x
- z} \right)} } \right)} ^2 , \label{pot2}
\end{equation}
where $ \left\langle H \right\rangle _0  = \left\langle 0
\right|H\left| 0 \right\rangle$. Following our earlier procedure
\cite{Gaete2}, we see that the second term on the right-hand side
of Eq. (\ref{pot2}) is clearly dependent on the distance, and the
potential for two opposite located at $\bf y$ and $\bf y \prime$
takes the form
\begin{equation}
V = \frac{{e^2 }}{{\pi }}\ln |\bf y - \bf y^ \prime| .
\label{pot3}
\end{equation}
Now we recall that the same calculation for the
Maxwell-Chern-Simons theory \cite{Gaete2,Elcio} gives
\begin{equation}
V =  - \frac{{e^2 }}{\pi }K_0 \left( \mu |\bf y - \bf y^ \prime|
\right), \label{potz}
\end{equation}
where $K_0$ is a modified Bessel function. We immediately see that
the result (\ref{pot3}) agrees with the behavior of the
Maxwell-Chern-Simons theory in the limit of short separation.
Eq.(\ref{pot3}) displays the confining nature of the potential
(the potential grows to infinity when the mutual separation
grows), but the Maxwell-Chern-Simons theory result (\ref{potz})
does not. In summary, the above analysis reveals that, although
both theories lead to fractional statistics by the same mechanism
of attaching a magnetic flux to the charges, the physical content
is quite different. However, the observation in the present work
that the new electrodynamics is confining is new.

It is worth noting here that there is an alternative but
equivalent way of obtaining the result (\ref{pot3}). To do this we
consider
\begin{equation}
V \equiv e\left( {{\cal A}_0 \left( \bf y \right) - {\cal A}_0
\left( {\bf y^ \prime} \right)} \right), \label{pot4}
\end{equation}
where the physical scalar potential is given by
\begin{equation}
{\cal A}_0 \left( {t,\bf x} \right) = \int_0^1 {d\lambda } x^i E_i
\left( {t,\lambda \bf x} \right). \label{pot5}
\end{equation}
Two remarks are pertinent at this point. First, Eq.(\ref{pot5})
follows from the vector gauge-invariant field \cite{Gaete}
\begin{equation}
{\cal A}_\mu  \left( x \right) \equiv A_\mu  \left( x \right) +
\partial _\mu \left( { - \int_\xi ^x {dz^\mu  } A_\mu  \left( z
\right)} \right), \label{pot6}
\end{equation}
where, as in Eq.(\ref{gen5}), the line integral appearing in the
above expression is along a spacelike path from the point $\xi$ to
$x$, on a fixed time slice. Second, it should be noted that the
gauge-invariant variables (\ref{pot6}) commute with the sole first
class constraint (Gauss' law), corroborating the fact that these
fields are physical variables \cite{Dirac}.

Having made these observations and from Eq. (\ref{ecua2}), we can
write immediately the following expression for the physical scalar
potential
\begin{equation}
{\cal A}_0 \left( {t,\bf x} \right) = \int_0^1 {d\lambda } x^i E_i
\left( {t,\lambda \bf x} \right) = \int_0^1 {d\lambda } x^i
\partial _i \left( {\frac{{ - J^0 \left( {\lambda \bf \bf x}
\right)}}{{\nabla _{\lambda \bf x}^2 }}} \right), \label{pot7}
\end{equation}
where $J^0$ is the external current. The static current describing
two opposite charges $e$ and $-e$ located at $\bf y$ and $\bf y
\prime$ is then given by $J^0 \left( {t,\bf x} \right) = e\left\{
{\delta ^{\left( 2 \right)} \left( {\bf x - \bf y} \right) -
\delta ^{\left( 2 \right)} \left( {\bf x - \bf y^ \prime }
\right)} \right\}$. Substituting this back into Eq. (\ref {pot7})
we obtain
\begin{equation}
V = \frac{{e^2 }}{\pi }\ln |\bf y - \bf y^ \prime|. \label {pot8}
\end{equation}
It is clear from this discussion that a correct identification of
physical degrees of freedom is a key feature for understanding the
physics hidden in gauge theories. According to this viewpoint,
once that identification is made, the computation of the potential
is achieved by means of Gauss law \cite{Haagensen}. As a final
comment, we point out that the methodology advocated in this paper
provides yet another support to the dressed fields picture to
compute the static potential in gauge theories.

\section{ACKNOWLEDGMENTS}

I would like to thank G. Cvetic for helpful comments on the
manuscript. I would also like to thank I. Schmidt for his support.

\end{document}